\documentclass[aps,prl,preprint,superscriptaddress]{revtex4-1}
\usepackage{graphicx}
\usepackage{gensymb}
\begin{document}


\title{Combined study of microwave-power- and linear-polarization- dependence of the microwave radiation-induced magnetoresistance oscillations in GaAs/AlGaAs devices}


\author{Tianyu Ye}
\author{Han-Chun Liu}

\affiliation{Department of Physics and Astronomy, Georgia State University, Atlanta, Georgia 30303, USA}

\author{W. Wegscheider}
\affiliation{Laboratorium f\"ur Festk\"orperphysik, ETH Z\"urich, 8093 Z\"urich, Switzerland}
\author{R. G. Mani}
\affiliation{Department of Physics and Astronomy, Georgia State University, Atlanta, Georgia 30303, USA}

\date{\today}

\begin{abstract}
We report the results of a combined microwave polarization-dependence- and power- dependence study of the microwave radiation-induced magnetoresistance oscillations in high mobility GaAs/AlGaAs heterostructure devices at liquid helium temperatures. The diagonal resistance was measured with the magnetic field fixed at the extrema of the  radiation-induced magnetoresistance oscillations, as the microwave power was varied at a number of microwave polarization angles. The results indicate a non-linear relation between the oscillatory peak- or valley- magnetoresistance and the microwave power, as well as a cosine square relation between the oscillatory peak- or valley- magnetoresistance and the microwave polarization angle. A simple model is provided to convey our understanding of the observations.
\end{abstract}


\maketitle

\section{introduction}
High mobility two dimensional electron systems (2DES) show extraordinary physical properties such as the Integral- and Fractional- Quantum Hall Effects at low temperatures and low filling factors, where the diagonal resistance vanishes as the  Hall resistance exhibits quantized Hall plateaus. More recently, it has been demonstrated that such high mobility 2DES can also exhibit zero-resistance states at very large filling factors or small magnetic fields when the 2DES material is photo-excited by microwave or terahertz radiation. Such microwave radiation-induced zero resistance states (MRiZRS),\cite{Maninature2002} arise from  large amplitude $1/B$-periodic microwave radiation-induced magnetoresistance oscillations (MRiMOs)\cite{Maninature2002,ZudovPRLDissipationless2003}. Although the observations reported thus far have helped to stimulate much theory for associated steady-state non-equilibrium transport in low dimensional electronic systems, there remain unsettled experimental issues that require resolution in order to further the understanding of transport in the 2DES under microwave excitation.

After nearly a decade of study, the issue of the phase of the MRiMOs has apparently been settled in favor of the ``1/4-cycle-shifted'' oscillations\cite{Maninature2002,ManiPRLPhaseshift2004} where the oscillatory minima occur about  $B=[4/(4j+1)]B_f$, i.e., for the $\omega = (j+1/4) \omega_{c}$  condition instead of for the $\omega=(j+1/2) \omega_{c}$ condition\cite{ZudovPRLDissipationless2003}. Here $B_f=2\pi fm^*/e$, $f$ is microwave frequency, $\omega = 2 \pi f$, $m^*$ is effective mass of electron in GaAs, $e$ is electron charge, $\omega_{c}$ is the cyclotron angular frequency, and $j = 1, 2, 3...$. However, there still remain two issues that remain unsettled: the power dependence of the amplitude of the microwave radiation-induced magnetoresistance oscillations and the linear polarization sensitivity of these same oscillations. The experimental\cite{ManiAPL2008,ManiPRBVI2004,ManiPRLPhaseshift2004,ManiEP2DS152004,KovalevSolidSCommNod2004,SimovicPRBDensity2005,ManiPRBTilteB2005,SmetPRLCircularPolar2005,WiedmannPRBInterference2008,DennisKonoPRLConductanceOsc2009,ManiPRBPhaseStudy2009,ManiPRBAmplitude2010,ArunaPRBeHeating2011,ManiPRBPolarization2011,ManinatureComm2012,ManiPRBterahertz2013,TYe2013,Mani2013sizematter, ManiNegRes2013} and
theoretical\cite{DurstPRLDisplacement2003,AndreevPRLZeroDC2003,RyzhiiJPCMNonlinear2003,KoulakovPRBNonpara2003,LeiPRLBalanceF2003,DmitrievPRBMIMO2005,LeiPRBAbsorption+heating2005,InarreaPRLeJump2005,ChepelianskiiEPJB2007,FinklerHalperinPRB2009,ChepelianskiiPRBedgetrans2009,InarreaPRBPower2010,MikhailovPRBponderomotive2011,Inarrea2011,Lei2012Polar,Inarrea2013Polar, Kunold2013, Zhirov2013} status of these aspects are as follows: For the power dependence, some experimental results\cite{ManiPRBAmplitude2010} have indicated that MRiMOs’ amplitude increase non-linearly with the microwave power, as the radiation driven electron orbit model\cite{InarreaPRBPower2010} has theoretically confirmed a non-linear power relation. In contrast, the  inelastic model\cite{DmitrievPRBMIMO2005} suggests that the amplitude of the radiation-induced magnetoresistance oscillations should increase linearly with the microwave power, as suggested in early experimental work[ \cite{ZudovPRLDissipationless2003}. So far as the microwave polarization dependence is concerned, one set of experiments\cite{SmetPRLCircularPolar2005} suggests that MRiMOs are independent to the polarization orientation for both linearly and circularly polarized microwaves, while another set of experiments\cite{ManiPRBPolarization2011} shows MRiMO-amplitudes do depend on the polarization angle of linear polarized microwave and that it follows a cosine square function rule. Theoretically, both a displacement model\cite{Lei2012Polar} and the radiation driven electron orbital model\cite{Inarrea2013Polar} confirm the dependence of MRiMOs on the polarization angle of linear polarized microwaves. Additionally, a displacement model\cite{LeiPRBAbsorption+heating2005} also indicated the  dependence of the oscillatory magnetoresistance on circular- and linear- polarization. However, the inelastic model\cite{DmitrievPRBMIMO2005} has strongly supported the idea that  MRiMOs are insensitive to the polarization for both linearly- and circularly- polarized microwaves.

\section{experiment and results}
To further examine the power- and linear polarization angle- dependence, we have carried out a combined study of microwave power- and linear polarization rotation angle- dependence of the amplitude of the MRiMOs. For such experiments, Hall bars with gold-germanium alloyed contacts were fabricated on high mobility GaAs/AlGaAs heterojunctions by optical lithography. The specimens were mounted at the end of a long cylindrical waveguide, with the device-normal oriented along the waveguide axis. The waveguide sample holder was then inserted into a variable temperature insert, inside the bore of a superconducting solenoid. A base temperature of approximately 1.5 K was realized by pumping on the liquid helium within the variable temperature insert. The specimens were briefly illuminated by a red LED light at low temperature to realize the high mobility condition. Finally, a low frequency four terminal lock-in technique was adopted to measure the magnetoresistance. A commercially available microwave synthesizer provided the microwave excitation and the microwave power at the source was changed at 1 dBm increments for the power dependence measurements. The linear polarization angle, which is defined as the angle between the long axis of the Hall bar and the microwave antenna in the microwave launcher, see figure 2(a), was changed by rotating the microwave launcher outside the cryostat. The results to be reported here are characteristic of the power- and polarization-angle-dependence  over the range $30 \le f\le 50$ GHz. However, we focus here on the results at just one frequency, 33.62 GHz, since so many plots need to be exhibited to establish the overall behavior even at one frequency.

\begin{figure}[t]
\centering
\includegraphics[width= 75mm]{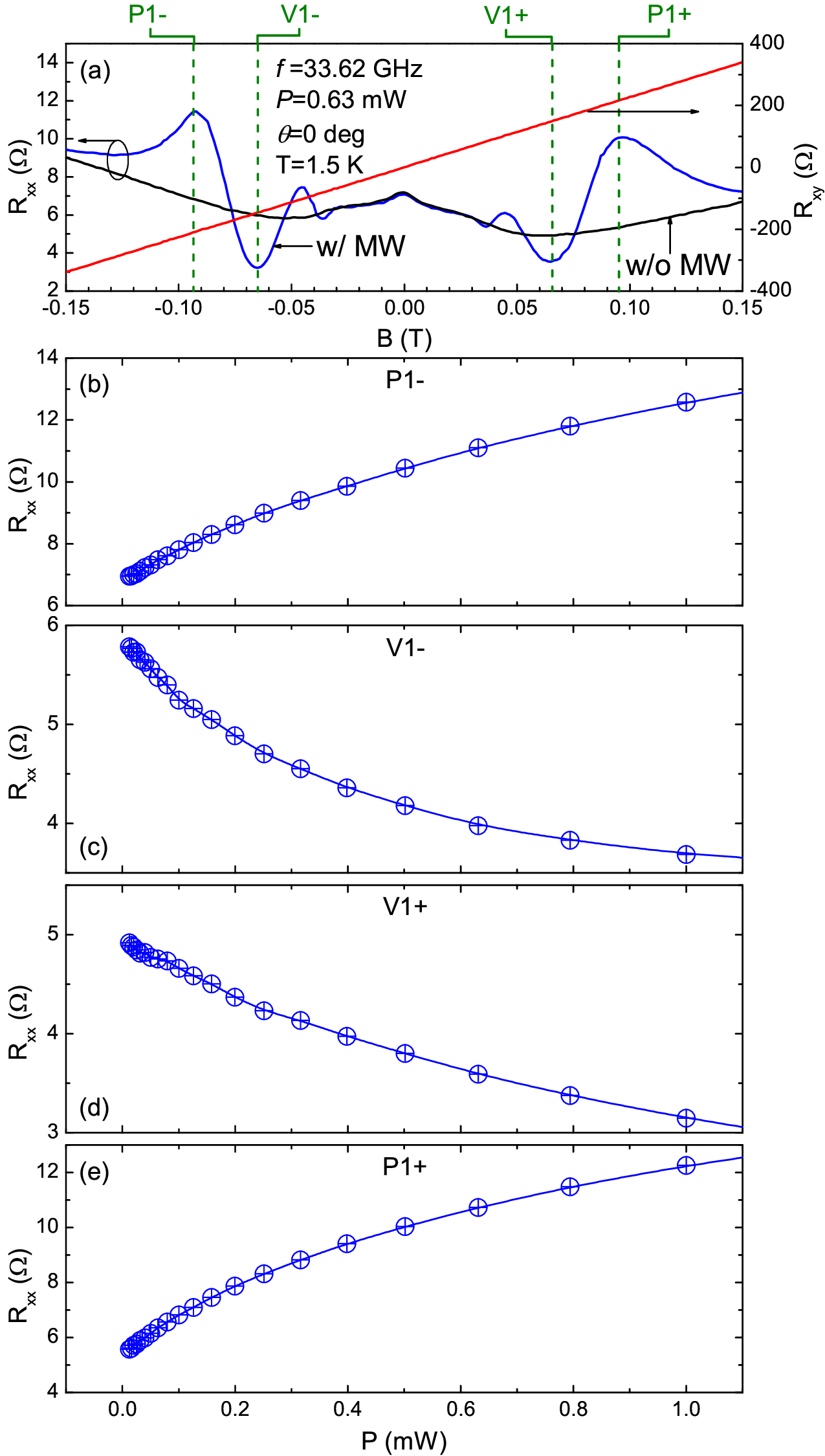}
\caption{(Color online) (a)Diagonal resistance $R_{xx}$ (left ordinate) and Hall resistance (right ordinate) versus the magnetic field $B$ without (black curve) and with (red- and blue- curves) microwave photo-excitation at 33.62 GHz and $T=1.5$ K. The polarization angle, $\theta $, is zero. Symbols in green at the top abscissa  mark the magnetic fields of some of the peaks and valleys of the oscillatory magnetoresistance. Panels (b)- (e) show the $R_{xx}$ as a function of the microwave power, $P$, at  these peaks and valleys as follows: (b) P1-, (c) V1-, (d) V1+ and (e) P1+.}
\end{figure}

A magnetic field sweep was performed with 33.62 GHz microwave illumination, see figure 1(a), to obtain the photo-excited diagonal magnetoresistance, $R_{xx}$, following a field sweep to obtain the dark $R_{xx}$ curve. The photo-excited blue $R_{xx}$ curve shows pronounced radiation-induced magnetoresistance oscillations on both sides of magnetic field axis. Since the peaks labeled P1- and P1+, and the valleys labeled V1- and V1+  deviate the most from the dark curve, we shall examine the  power dependence- and polarization dependence-  at the associated four fixed values of the magnetic field. Figure 1 (b) to (e) exhibit the extremal oscillatory diagonal magnetoresistance $R_{xx}$ as a function of the microwave power, in units of mW, for P1-, V1-, V1+ and P1+, respectively. Since Figure 1(b) and 1(e) exhibit the power-dependence at the peaks of the oscillatory resistance, the $R_{xx}$ increases as the power increases.  Correspondingly, since Fig. 1(c) and 1(d) show the power-dependence at the valleys of the oscillatory resistance, the $R_{xx}$ becomes smaller as the power increases. Note that neither the increase of $R_{xx}$ with $P$ in Fig. 1(b) and 1(e), nor the decrease of $R_{xx}$ with $P$ in Fig. 1(c) and 1(d)  is linear with respect to the microwave power. Instead, they appear to follow a power function: $R_{xx}(P)=a+c(P)^{\alpha}$, where a, c and $\alpha$ are parameters. A fit of the $R_{xx}$ vs $P$ traces indicated that the  $\alpha$ values for the different curves are all  $\approx$ 0.5 ($\pm$0.1), which is in agreement with the results reported in ref.\cite{ManiPRBAmplitude2010} and ref.\cite{InarreaPRBPower2010}.

\begin{figure}[t]
\centering
\includegraphics[width= 75mm]{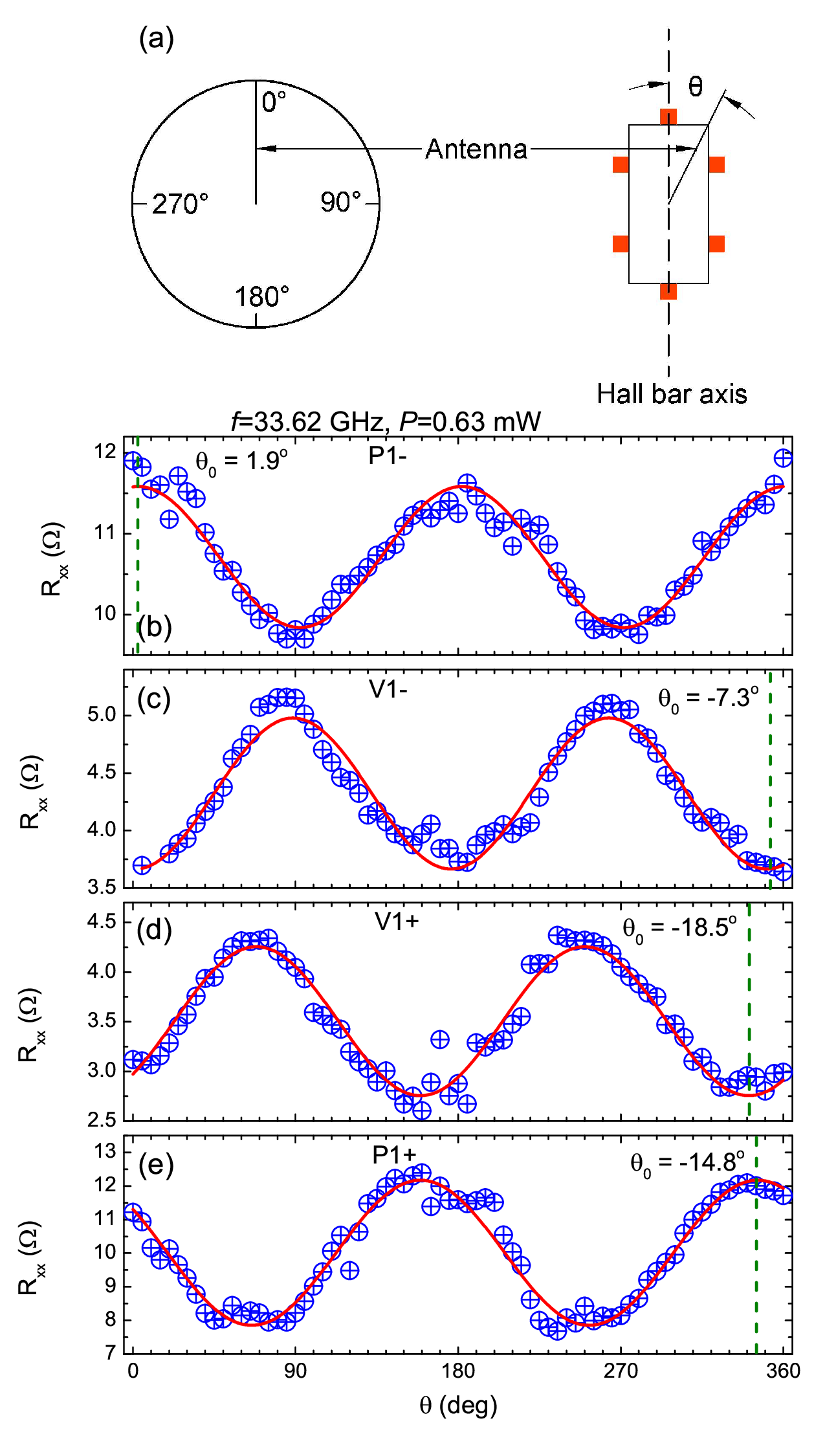}
\caption{(Color online) (a)A sketch of the polarization orientation in the magnetotransport measurement. Here, the antenna and the microwave launcher rotate clockwise with respect of the long axis of Hall bar sample to set the polarization angle $\theta$. (b) to (e) $R_{xx}$ versus the polarization angle $\theta$ at the magnetic fields corresponding to the oscillatory extrema ( P1-, V1-, V1+, and P1+) marked in figure 1(a). Here, microwave frequency is 33.62 GHz and power is 0.63 mW. Red lines are the fit curves to  $R_{xx}(\theta)=A\pm Ccos^2(\theta-\theta_0)$. Here, the ``+'' sign applies for (b) and (e), while the ``-'' sign applies for (c) and (d). Vertical dashed lines mark the polarization phase shift angle, $\theta_{0}$, for each fit curve.}
\end{figure}

Such measurements of $R_{xx}$ vs. $P$ at the magnetic fields corresponding to P1-, V1-, V1+, and P1+ were carried out at a number of linear polarization angles over the range $0 \degree \le \theta \le 360 \degree$ at $10 \degree$ increments. Figure 2 summarizes the extracted angular dependence at a fixed power, $P$= 0.63 mW. Here, Fig. 2(a) sketches the convention for reporting the linear polarization angle, while Fig. 2(b) - Fig. 2(e) show $R_{xx}$ vs. $\theta$ for P1-, V1-, V1+, and P1+, respectively. From Fig. 2(b) -2(e), it is clear that all $R_{xx}$ vs. $\theta$ traces vary sinusoidally with $\theta$. Indeed, the data curves follow the fit function\cite{ManiPRBPolarization2011} $R_{xx}(\theta)=A\pm Ccos^2(\theta-\theta_0)$ shown in red in the figures. The fit results confirmed that the data of Fig. 2(b) - 2(e) all showed a period of $\pi$, as expected. Further, the phase shift $\theta_{0}$ = $1.9 \degree$ in Fig. 2(b), $\theta_{0}$ = $-7.3 \degree$ in Fig. 2(c), $\theta_{0}$ = $-18.5 \degree$ in Fig. 2(d), and $\theta_{0}$ = $-14.8 \degree$ in Fig. 2(e). These extracted phase shifts indicate a non-trivial phase shift under field reversal, i.e., the phase changes by $16.7 \degree$ from P1- to P1+, and $11.2 \degree$  from V1- to V1+.

\begin{figure}[t]
\centering
\includegraphics[width= 75mm]{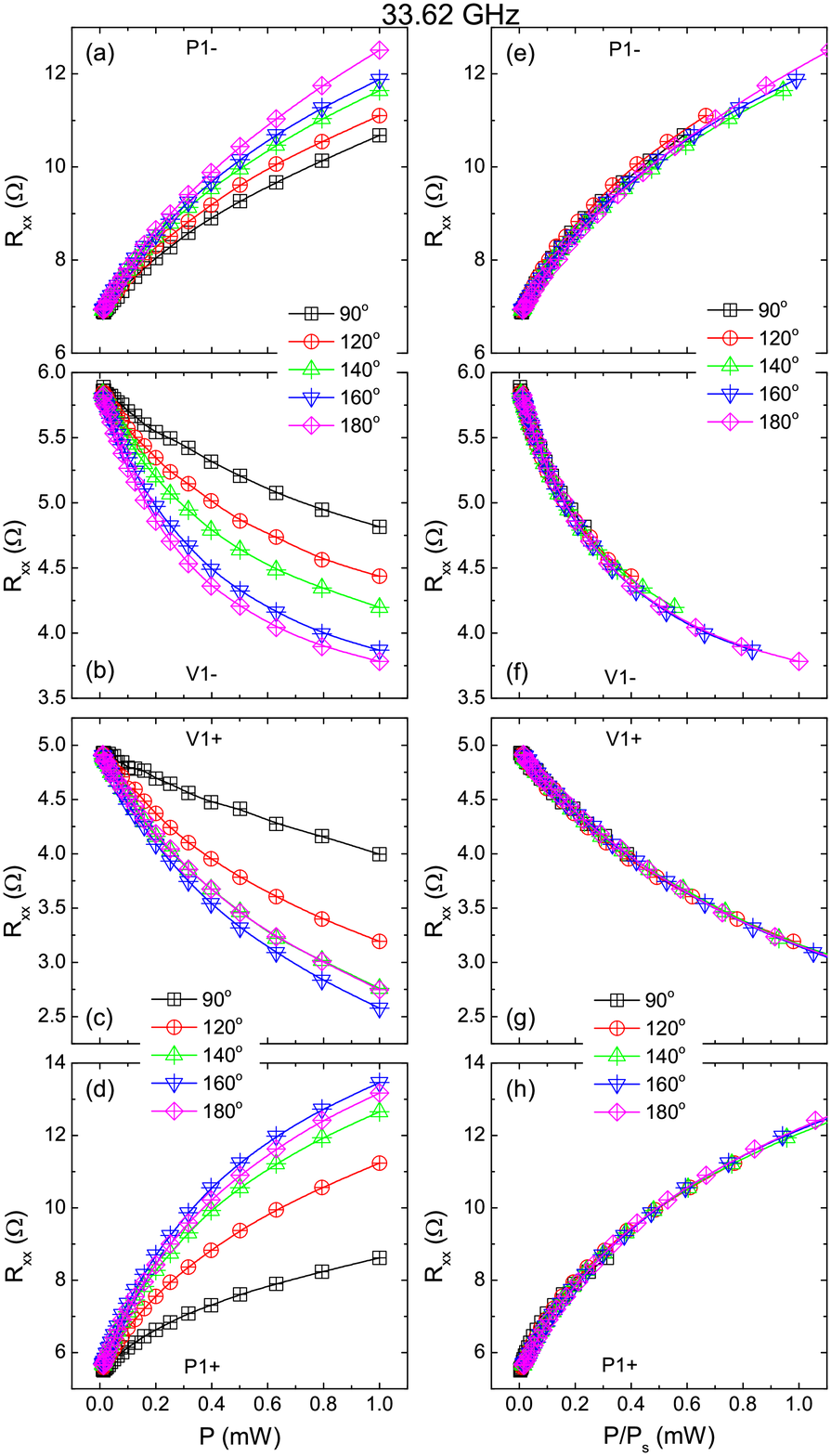}
\caption{(Color online) In plots (a) - (d), the diagonal resistance $R_{xx}$ is plotted versus the microwave power $P$ at various fixed polarization angles for the $R_{xx}$ extrema as follows: (a)P1-, (b)V1-, (c)V1+  and (d)P1+. In these plots, the different symbols  stand for different polarization angles from $90 \degree$ to $180 \degree$, as indicated. Plots (e) to (h) show the extremal $R_{xx}$ versus $P/P_s$. Here, $P_s$ is a polarization angle dependent power scaling factor that normalizes curves on the left column at different polarization angles to the same curve. }
\end{figure}

Figure 3 shows the  $R_{xx}$ vs $P$ obtained for different polarization angles at P1-(Fig. 3 (a)), V1- (Fig. 3(b)), V1+ (Fig. 3(c)), and P1+ (Fig. 3(d)). For the sake of clarity, each of these plots shows only a few representative polarization angles in the range $90 \degree \le \theta \le 180 \degree$. Although a non-linearity in the $R_{xx}$ vs. $P$ curves is apparent for all the curves, the change in $R_{xx}$ between $P=0$ and $P=1$ mW is not the same at all polarization angles.  A close examination suggests, however, that all the curves in a given panel follow the same power function law. Therefore, we utilize a power scaling factor $P_s$ to normalize the different polarization angle curves in a given panel (Fig. 3(a)-(d)) to the same curve. As figure 3 (e) to (h) show, all the curves on the left column of Fig. 3 can be normalized to the same curve on the right column by using a polarization angle dependent power scaling factor, i.e., $P_{s} (\theta)$. These power scaling factors depend upon both the magnetic field and polarization angles. We have found that the  power scaling factor follows a certain rule as the polarization angle change, see Fig. 4.

\begin{figure}[t]
\centering
\includegraphics[width= 75mm]{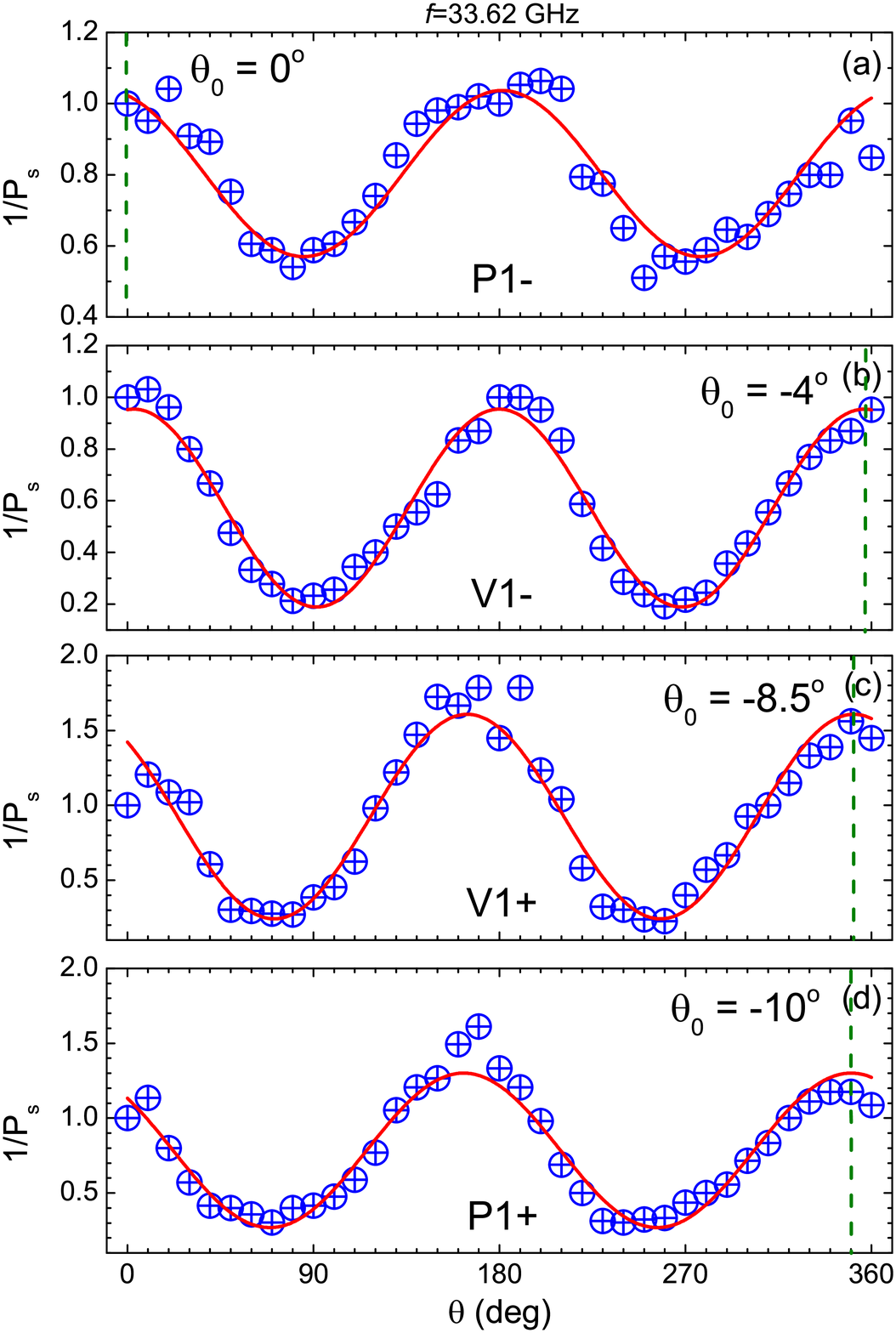}
\caption{(Color online) (a) The inverse power scaling factor, $1/P_{s}$, is plotted as a function of linear microwave polarization angle $\theta$ at the oscillatory magnetoresistance extrema: (a)P1-, (b)V1-, (c)V1+ and (d)P1+. Red curves represent the fit utilizing $1/P_{s} (\theta)=A+Ccos^2(\theta-\theta_0)$. Vertical dash lines indicate the polarization phase shift angle, $\theta_{0}$, obtained from the fit.}
\end{figure}

In figure 4, the inverted power scaling factor $1/P_s$ is plotted versus the polarization angle $\theta$  for P1- (Fig. 4(a)), V1-  (Fig. 4(b)), V1+ (Fig. 4(c)), and P1+ (Fig. 4 (d)). Fig. 4 shows that the $1/P_s$ varies sinusoidally with the polarization angle $\theta$, and follows the fit function $R_{xx}(\theta)=A+Ccos^2(\theta-\theta_0)$, which is shown in red in the figures. The main difference between these fits and the fits shown in Fig. 2 is that, here, both the peaks and the valleys in the oscillatory magnetoresistance  show a ``+'' sign between the constant (A) and the $Ccos^{2}(\theta-\theta_0)$ terms, while for the data of Fig. 2, the ``+'' sign appears for oscillatory magnetoresistance maxima, while the ``-'' term appears for the oscillatory magnetoresistance minima.  As the Fig. 4(a) to (d) show,  the polarization phase shift angle $\theta_{0}$ is less than $10 \degree$ and comparable to the values reported in Fig. 2.  For instance, $\theta_0$ observed in the  $1/P_s(\theta)$ vs. $\theta$ plot is $0 \degree$ in Fig. 4(a) and that observed in the  $R_{xx}(\theta)$ vs. $\theta$ plot is $-1.9 \degree$ in Fig. 2(b). The  experimental uncertainty in associated polarization angle measurements is $10 \degree$.

\section{discussion}
From the above results, it is possible to draw three conclusions: (a) the magnetoresistance $R_{xx}$ is a non-linear function of the microwave power $P$, see Fig. 1 and Fig. 3.  (b) the peak- or valley- magnetoresistance is a cosine square function of linear microwave polarization angle $\theta$, i.e., $R_{xx}(\theta)=A\pm Ccos^2(\theta-\theta_0)$, see Fig. 2. (c) Although the $R_{xx}$ vs. $P$ traces at different polarization angles for a given extremum look dissimilar at first sight, see Fig. 3(a)-(d), they are really just manifestations of the same trace because the $R_{xx}$ vs $P$ curves for different polarization angles can be normalized  by dividing with a power scaling factor $P_s$, see Fig. 3(e) - (h). Remarkably, this scaling factor also follows a cosine square function of linear microwave polarization angle, see  Fig. 4.  Note that, although we have only exhibited the results for $f$= 33.62 GHz, the same features were observed at other frequencies in the range $30 \le f \le 50$ GHz.

The first two points show the relative success of the displacement model and the radiation driven electron orbital model in describing the power-dependence and polarization-dependence of the microwave radiation-induced magnetoresistance oscillations. The displacement model\cite{DurstPRLDisplacement2003,LeiPRLBalanceF2003} suggests that microwave photo-excited electrons are scattered by impurities, and this gives rise to an additional current density due to radiation. The magnitude of this photo-excited current density is a function of the polarization angle. However, to our knowledge, there is no clear prediction regarding how, quantitatively, the microwave power at different polarization angles influences the photo-excited current density. The radiation driven electron orbital model\cite{InarreaPRLeJump2005} describes a periodic back- and forth- radiation driven motion of the electron orbits and the  conductivity modulation resulting from the average coordinate change (scattering jump). This model shows both a non-linear power dependence and a linear polarization angle dependence, similar to experiment. On the other hand, the inelastic model for magnetooscillations in the photoconductivity of the 2DES\cite{DmitrievPRBMIMO2005}, is governed by the microwave-induced change in the distribution function. Here, steady state microwave photoexcitation produces a change in the distribution function with oscillatory components that lead to oscillatory variations in the photoconductivity, which are linear in the microwave power and independent of the linear microwave polarization.

Finally, regarding point (c), the point that the $R_{xx}$ vs. $P$ curves at different polarization angles $\theta$ are really just manifestations of the same curve can be understood as follows: In our experiment, the transverse electric (TE) mode is excited by the MW antenna of figure 2(a), and the specimen is subject to the TE$_{11}$ mode of the circular waveguide. The electric field along the device axis or the effective electric field $E_e$, see Fig. 2(a),  is $E_e=Ecos\theta$, where $E$ is applied AC electric field. Since the measurements are reported as a function of the microwave power, which is the experimental variable, one might define an effective power, $P_e\propto E_e^2$. Then, $P_e = P cos^2\theta$ gives the relation between the effective and applied microwave power and the polarization angle. The results, see Fig. 3,  show that $P/P_s$ normalizes microwave power with different polarization angles to the same effective power scale, making $R_{xx}$ vs $P/P_s$ curves overlap each other on the right column of figure 3. Further, figure 4 shows that $1/P_{s}$ vs. $\theta$ can be fit  $1/P_{s}(\theta)=A+Ccos^2(\theta-\theta_0)$. This confirms the role of a device parallel electric field. Fig. 4 also indicates that the phase shifts of this power scaling factor deviates at most by small angles from $0 \degree$, which suggests that the zero polarization angle yields the maximum effective power. 

\section{conclusion}
Microwave power dependence and linear polarization dependence studies of the radiation-induced oscillatory magnetoresistance in high quality GaAs/AlGaAs 2DES indicate a non-linear variation in the amplitude of the radiation-induced magnetoresistance oscillations with the microwave power at the oscillatory extrema along with a cosine squared dependence on the polarization angle.  Furthermore, an empirically defined power scaling factor for normalizing the $R_{xx}$ vs. $P$ traces obtained at different linear microwave polarization angles also follows the cosine square function. This latter feature suggests that the device parallel component of electric field influences photo-excited electron transport. However, because the phase shifts $\theta_{0}$  are different at different magnetic fields, there are most likely other factors other than the cosine component of electric field that also play a role in the observed response.

\section{acknowledgement}
Basic research and helium recovery at Georgia State University is supported by the U.S. Department of Energy, Office of Basic Energy Sciences, Material Sciences and Engineering Division under DE-SC0001762. Additional support is provided by the ARO under W911NF-07-01-015.

\pagebreak
\section*{Figure Captions}
Figure 1: (a)Diagonal resistance $R_{xx}$ (left ordinate) and Hall resistance (right ordinate) versus the magnetic field $B$ without (black curve) and with (red- and blue- curves) microwave photo-excitation at 33.62 GHz and $T=1.5$ K. The polarization angle, $\theta $, is zero. Symbols in green at the top abscissa  mark the magnetic fields of some of the peaks and valleys of the oscillatory magnetoresistance. Panels (b)- (e) show the $R_{xx}$ as a function of the microwave power, $P$, at  these peaks and valleys as follows: (b) P1-, (c) V1-, (d) V1+ and (e) P1+.

Figure 2: (a)A sketch of the polarization orientation in the magnetotransport measurement. Here, the antenna and the microwave launcher rotate clockwise with respect of the long axis of Hall bar sample to set the polarization angle $\theta$. (b) to (e) $R_{xx}$ versus the polarization angle $\theta$ at the magnetic fields corresponding to the oscillatory extrema ( P1-, V1-, V1+, and P1+) marked in figure 1(a). Here, microwave frequency is 33.62 GHz and power is 0.63 mW. Red lines are the fit curves to  $R_{xx}(\theta)=A\pm Ccos^2(\theta-\theta_0)$. Here, the ``+'' sign applies for (b) and (e), while the ``-'' sign applies for (c) and (d). Vertical dashed lines mark the polarization phase shift angle, $\theta_{0}$, for each fit curve.

Figure 3:  In plots (a) - (d), the diagonal resistance $R_{xx}$ is plotted versus the microwave power $P$ at various fixed polarization angles for the $R_{xx}$ extrema as follows: (a)P1-, (b)V1-, (c)V1+  and (d)P1+. In these plots, the different symbols  stand for different polarization angles from 90$\degree$ to 180$\degree$, as indicated. Plots (e) to (h) show the extremal $R_{xx}$ versus $P/P_s$. Here, $P_s$ is a polarization angle dependent power scaling factor that normalizes curves on the left column at different polarization angles to the same curve.

Figure 4: (a) The inverse power scaling factor, $1/P_{s}$, is plotted as a function of linear microwave polarization angle $\theta$ at the oscillatory magnetoresistance extrema: (a)P1-, (b)V1-, (c)V1+ and (d)P1+. Red curves represent the fit utilizing $1/P_{s} (\theta)=A+Ccos^2(\theta-\theta_0)$. Vertical dash lines indicate the polarization phase shift angle, $\theta_{0}$, obtained from the fit.

\pagebreak

\begin{figure}[t]
\centering
\includegraphics[width=85 mm]{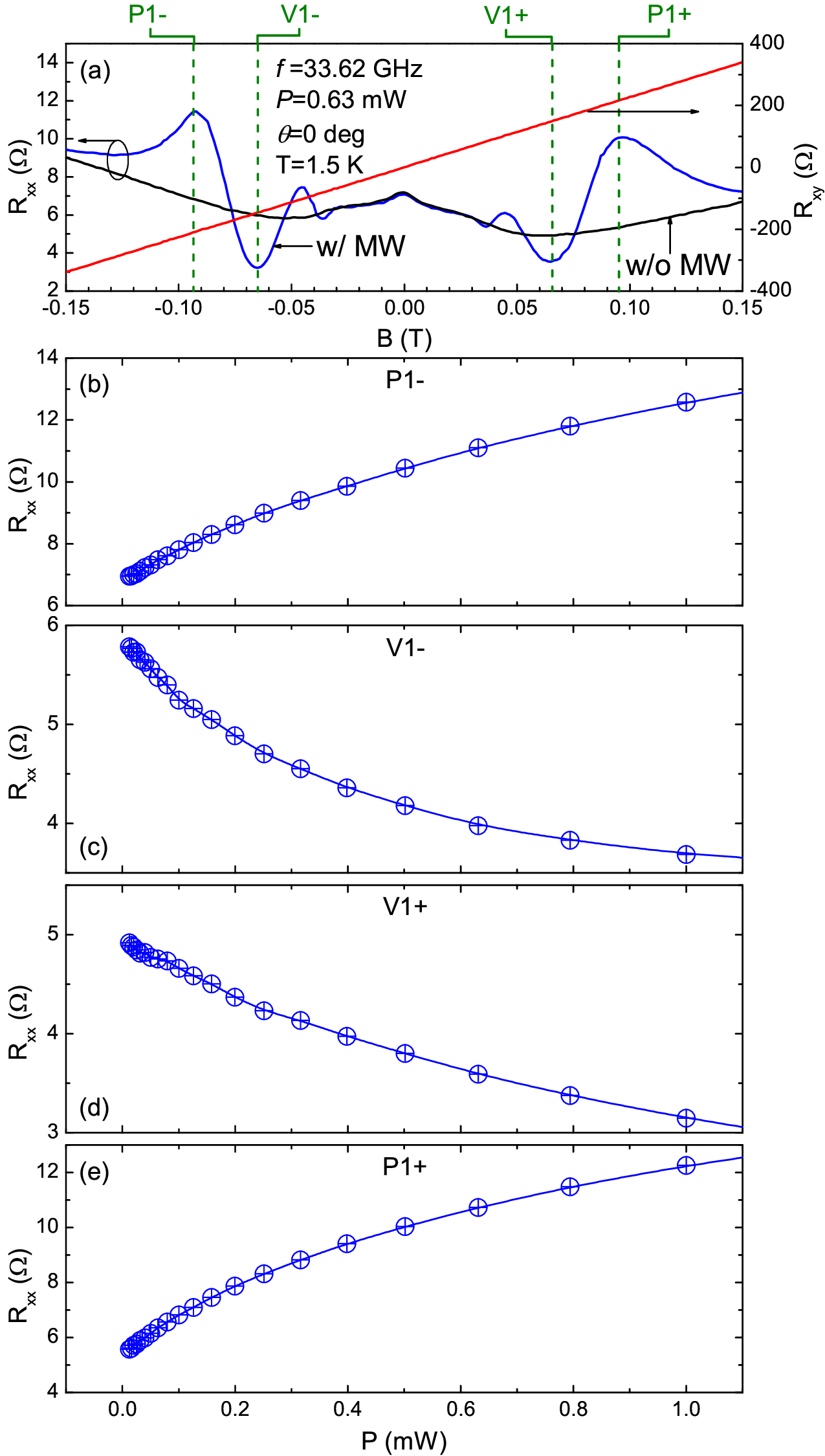}
\begin{center}
Figure 1
\end{center}
\end{figure}

\begin{figure}[t]
\centering
\includegraphics[width=85 mm]{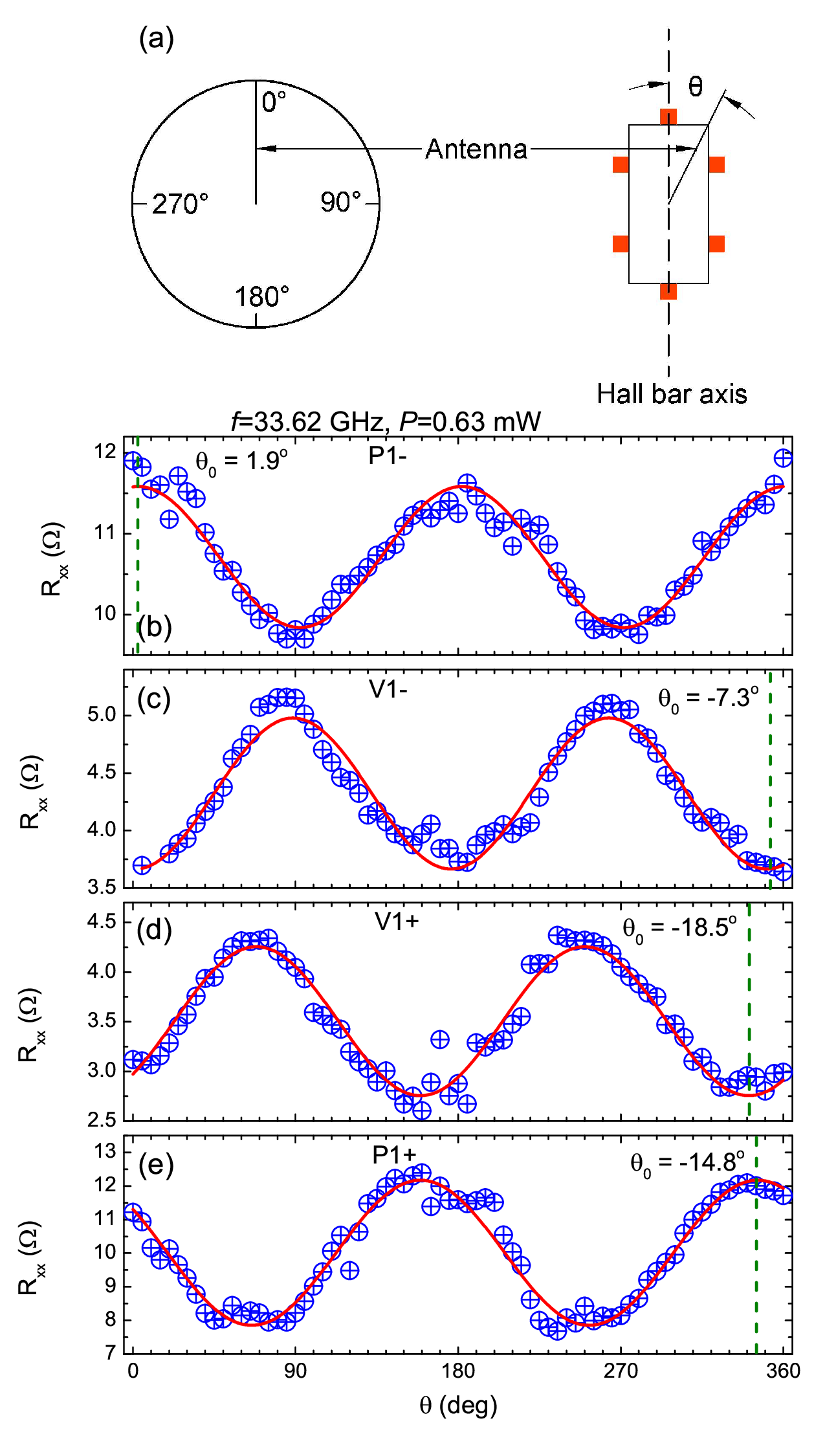}
\begin{center}
Figure 2
\end{center}
\end{figure}

\begin{figure}[t]
\centering
\includegraphics[width=85 mm]{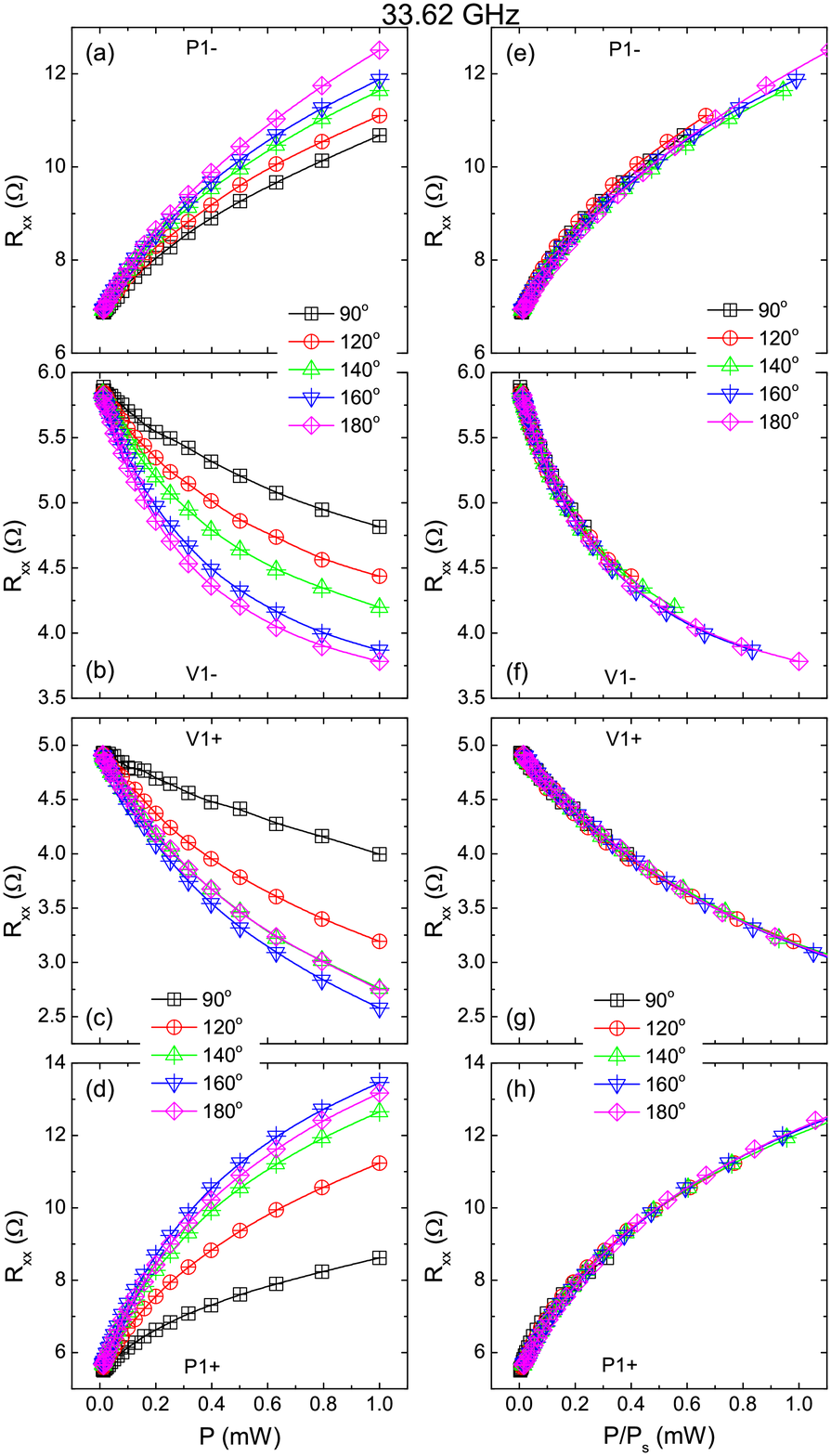}
\begin{center}
Figure 3
\end{center}
\end{figure}

\begin{figure}[t]
\centering
\includegraphics[width=85 mm]{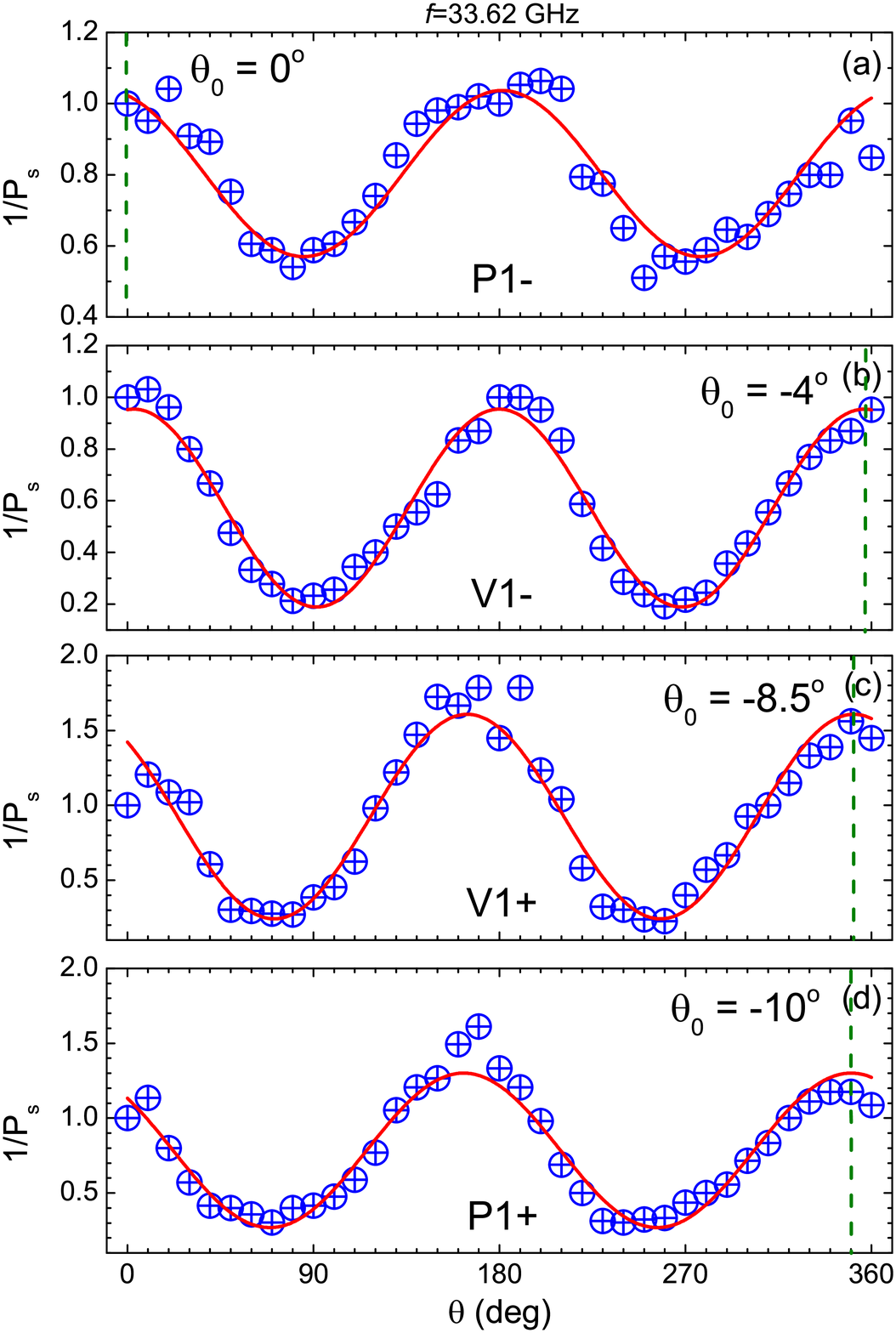}
\begin{center}
Figure 4
\end{center}
\end{figure}

\end{document}